# Flux Linkage Based Evaluation Method for Voltage Inertia and Voltage Recovery Capability Under Large Disturbances

***Abstract*—High-voltage direct current (HVDC) transmission applications and the growth of the dynamic load in large-scale receiving-end grids lead to a higher risk of short-term voltage instability. An effective way to address this problem is to improve the system's dynamic voltage support capability by changing the operation status of dynamic var devices, and the dynamic var reserve (DVR) is commonly used. Due to the time delay in synchronous machine excitation systems, the dynamic var reserved at steady state cannot be exploited immediately under large disturbances. In addition, some reactive power is produced immediately through electromagnetic induction. The voltage support effect of the two capabilities is analyzed based on the flux linkage and an approximate simulation of the fault impact. Then two novel indexes for evaluating the voltage inertia and voltage recovery capability are proposed, which are related to the voltage nadir and voltage recovery speed. The indexes' physical meanings are revealed by comparison with the frequency response, and the potential applications in planning and optimal reactive power dispatch (ORPD) are introduced. Numerical simulations based on the IEEE 39-bus system verify that the indexes can quantify the voltage support capabilities, and the minimum voltage support requirements are obtained to maintain systems' security.**

## NOMENCLATURE

### *A. Acronyms*

| | |
|---|---|
| DVR | Dynamic var reserve |
| DVC | Dynamic voltage support capability |
| HVDC | High-voltage direct current |
| ORPD | Optimal reactive power dispatch |
| STVS | Short-term voltage stability |
| UC | Unit commitment |

### *B. Sets and Parameters*

| | |
|---|---|
| $K_A/T_e$ | Gain/time constant of an exciter |
| $Q_{max}$ | Upper limit of the reactive power output of a generator |
| $R_2$ | Rotor resistance of an induction motor |
| $S_1 / S_2 / S_{flt}$ | Sets of secure operation points/faults |
| $T'_{d,0} / T'_d$ | Time constant of excitation winding (open-circuit/short-circuit) |
| $T'_0$ | Time constant of an induction motor rotor circuit |
| $x_d / x'_d / x_q$ | d-axis/d-axis transient/q-axis reactance of generators |
| $x_{ad}$ | Armature reaction reactance |
| $x_f$ | Magnetizing reactance |
| $X_1 / X_2$ | Stator/rotor leakage reactance of an induction motor |
| $Y$ | Network parameters |

### *C. Variables*

| | |
|---|---|
| $E_{fd,0}$ | Initial value of the excitation voltage |
| $E_{q,0}$ | No-load electromotive force |
| $N_{dyn} / N_g$ | Number of dynamic elements/generators |
| $P_m$ | Active power of an induction motor |
| $P_g / Q_g / V_g / \delta / w$ | Initial values of the active power/reactive power/voltage/power angle/angular velocity of generators |
| $s_0$ | Initial slip of an induction motor |
| $T_{clr}$ | Moment when a fault is cleared |
| $V / V_d / V_q$ | Voltage/d-axis/q-axis component |
| $V_{flt} / V_{d,flt} / V_{q,flt}$ | Voltage/d-axis/q-axis component when a fault occurs |
| $V_{LD,0}$ | Initial value of load bus voltage |
| $VIC / VRC$ | Voltage inertia/voltage recovery capability index |
| $VIR / VRR$ | Requirements of voltage inertia/voltage recovery capability |
| $\psi'_d / \Delta\psi'_d / \psi'_{d,spon}$ | d-axis flux linkage/variation/spontaneous component |
| $V_i / V_{ij}$ | voltage/component of generator *j* on bus *i* |
| $V_{clr}$ | Voltage when a fault is removed |

## I. INTRODUCTION

With the wide application of high-voltage direct current (HVDC), several large-scale receiving-end power grids have been formed in eastern and southern China. Receiving-end power grids are faced with the increasingly prominent problem of short-term voltage stability (STVS). First, due to the increase

.

in dynamic loads in receiving-end power grids and the power fed in from HVDC, a large capacity of reactive power will be absorbed under a fault disturbance, which has a considerable impact on system voltage recovery[1]. In addition, to achieve DC power consumption and meet the requirements of peak regulation, a large number of conventional power resources, such as synchronous generators, are shut down, resulting in insufficient dynamic voltage support capability (DVC) in power systems. To ensure the STVS of receiving-end power grids, countermeasures can be taken, including preventive control[2-3], emergency control[4-7], planning[8-9], etc. The objective of preventive control is to evaluate the STVS in advance[10-12]. When a potential security hazard is detected under some faults, the operation points[2] or the on-off statuses[3] of reactive power devices are adjusted to ensure the systems' STVS after faults occur.

Compared with reactive power compensation devices based on power electronics such as SVC and STATCOM, synchronous generators and condensers can provide systems with short-circuit capacity and momentum of inertia [13]. Furthermore, they can provide a large capacity of reactive power to support system voltage recovery when faults occur [14]. Hence, these devices play an essential role in suppressing DC commutation failure and improving systems' STVS. In preventive control, by optimizing the operation states of these devices under steady state, the voltage recovery performance of the power grids can be economically improved. Previous studies generally used the difference between the upper limit and the current var output of dynamic var devices, defined as the dynamic var reserve (DVR), to search for an operating point that is better for system voltage recovery. Due to the differences in the installation locations of these devices, the DVR of the system is not a summation of all devices' DVR, and the difference in the voltage support effect of various devices should be considered. When evaluating the support effect of the DVR, methods based on statistical analysis [15-16] and sensitivity analysis [17-26] are commonly adopted. The objective of the former is to fit the relationship between the DVR and the voltage stability margin in linear [15] or high-order polynomial [16] form. The latter method calculates the sensitivity of reactive power to bus voltage based on the power flow Jacobian matrix [17] or the trajectory sensitivity [18-19]. In addition, the DVC can be calculated from the perspective of reactive power current injection [20-26]. When calculating the DVR capacity, the upper limit can be defined based on the rated reactive power of generators [15-18], the capacity of generators [16], or other parameters.

It is believed that the more DVR reserved under steady state, the more dynamic var support the generators/condensers can provide under large disturbances. However, due to the time delay in the excitation systems, it takes a few seconds for the devices to fully utilize the dynamic var reserved under steady state during this transient process. This is defined as the excitation control var response in this paper, which determines the voltage recovery performance. In addition, the flux linkage of the synchronous machine windings cannot change suddenly when a fault occurs. Thus, some current is induced in the windings through electromagnetic induction to keep the flux linkage unchanged, and the reactive power is also induced, which is defined as the spontaneous var response in this paper. Note that this response is not generated by the excitation systems' adjustment, and it can be generated immediately after a fault occurs, which is closely related to the voltage sag.

Since the capacity of the two kinds of var output is related to the devices' initial reactive power, voltage setting, unit parameters, etc., it is improper to evaluate the DVC just from the perspective of the steady-state var output. Therefore, two novel indexes considering different state variables are defined in this paper. When optimizing these indexes, coordination among different operation variables under steady state can be realized, and the voltage support ability under large disturbances can be fully exploited. In addition, the var output of the devices undergoes abrupt changes under large disturbances, while the flux linkage inside the machines does not change suddenly. The flux linkage supports the bus voltage by cutting the windings at a nearly constant speed, and its initial value will determine the initial voltage sag depth under a specific fault. For transmission systems, due to the action of fast var devices and the short duration of faults, the initial voltage sag determines the nadir dominantly, and the voltage changing rate is affected by the action of excitation systems and other devices, which are not the inherent support ability. Since the moment of inertia determines the frequency nadir dominantly [27] and is quantified by rotors' kinetic energy, we call the magnetic field energy storage or the flux linkage as voltage inertia. It is different from the cases in microgrids with high penetration of power electronic devices [28-29]. To sum up, the subproblems and the corresponding solutions are listed as follows.

(1) How can the spontaneous var response and excitation control var response be calculated? What is the relationship between the steady-state operation status and these var responses? Since the spontaneous var response is induced due to flux linkage conservation, we use this conservation as a condition for calculation in section II. For the excitation control var response, some simplification is adopted to obtain an analytical expression. The impact of devices' steady-state status on the dynamic var support capability under large disturbances is also investigated from the perspective of flux linkage.

(2) How can the voltage support capability of different devices' dynamic var be calculated based on their dynamic characteristics and the coupling levels among different buses? Since both the spontaneous var response and the excitation control var response are achieved by acting on the flux linkage, the change in the flux linkage can reflect the change in voltage support effect. Therefore, in section III we derive the time-domain expression of a generator's voltage support effect based on the generator's flux linkage expression in section II and the network equations. With this expression, two novel indexes for measuring the voltage inertia and voltage recovery ability under large disturbances are proposed. In addition, the physical meanings of the indexes and the functional orientation of the two kinds of capabilities are investigated through comparison with the frequency response.

(3) For a multimachine system, can the voltage support capability of the system maintain STVS under different fault disturbances? In section IV, we evaluate the requirements of voltage inertia and voltage recovery ability for the system under different faults based on the indexes proposed in section III. The evaluation can be conducted before planning or optimal reactive power dispatch (ORPD), and the results can be used as security constraints to add to planning or ORPD models.

The main contributions of this paper are listed as follows.

(1) It is revealed that the voltage support capability can be divided into two parts: the voltage inertia and voltage recovery capability. They are closely related to the initial level and changing rate of the magnetic field energy, which is the essence of voltage support capability and cannot change suddenly.

(2) With a high penetration of renewable energy in power systems, the definition of reactive power will be challenged since it is defined under sinusoidal conditions. The indexes in this paper are defined based on the flux linkage. It provides a novel viewpoint for analyzing the voltage support capability, which can also be used under non-sinusoidal conditions and thus is more universal.

(3) According to the indexes, the voltage support capability of the system in the current state and the voltage support requirements under different faults for maintaining the systems' security can be evaluated, which can serve as constraints in the models of planning and ORPD.

The remainder of this paper is organized as follows. In section II, the composition of the var response is investigated. In section III, two indexes are proposed to measure the voltage support effect of a generator to a bus from the flux linkage. The potential applications of the indexes are introduced in section IV. section V verifies the correctness and effectiveness of the analysis methods and the indexes through numerical simulations. The conclusions are drawn in section VI.

## II. MECHANISM ANALYSIS OF THE REACTIVE POWER RESPONSE

Under large disturbances, variables such as bus voltage and generator reactive power output in the power system undergo sudden changes. Due to flux linkage conservation when the topology changes, the generators' winding flux linkage remains unchanged at the moment of fault occurrence and removal. After that, the flux linkage changes under the control of excitation systems. Thus, the flux linkage can be used to analyze the changes in other variables at different stages of the transient process. Since this paper focuses on analyzing the mechanism of rotating reactive power sources' support effect on bus voltage, the impact of the DC system on the AC system voltage is ignored in this paper, which has been discussed in other studies [30-31]. In addition, since the mathematical model of synchronous generators is similar to that of synchronous condensers, we take only the former as an example due to space limitations.

### *A. Approximate time-domain expression of the generator winding flux linkage*

The third-order model of a synchronous generator is utilized when deriving the analytical expression of the flux linkage. In fact, the following analysis method can also be adopted for other order models. Due to the inertia of generator rotors and the relatively short time span considered, the impact of the power angle change on the bus voltage is much smaller than that of the dynamic load and the excitation system. Therefore, it can be ignored. For the excitation system, the model in [32] is adopted. However, to solve this model, it is necessary to use numerical methods. To obtain an approximate analytical expression for analysis, we assume that the terminal voltage of the generators remains unchanged during the fault period and in a short period after the fault is removed. Thus, the curve displays a 'stepped' shape, as shown by the red line in Fig 1. $V_{\text{flt}}$ and $V_{\text{clr}}$ can be calculated based on the network topology and the value of the flux linkage at the moment of fault occurrence and removal, as shown in section III. In this way, the impact of the fault on generators' transient response can be simulated. It should be noted that to improve the accuracy, the transient process can also be divided into additional periods to obtain improved approximations.

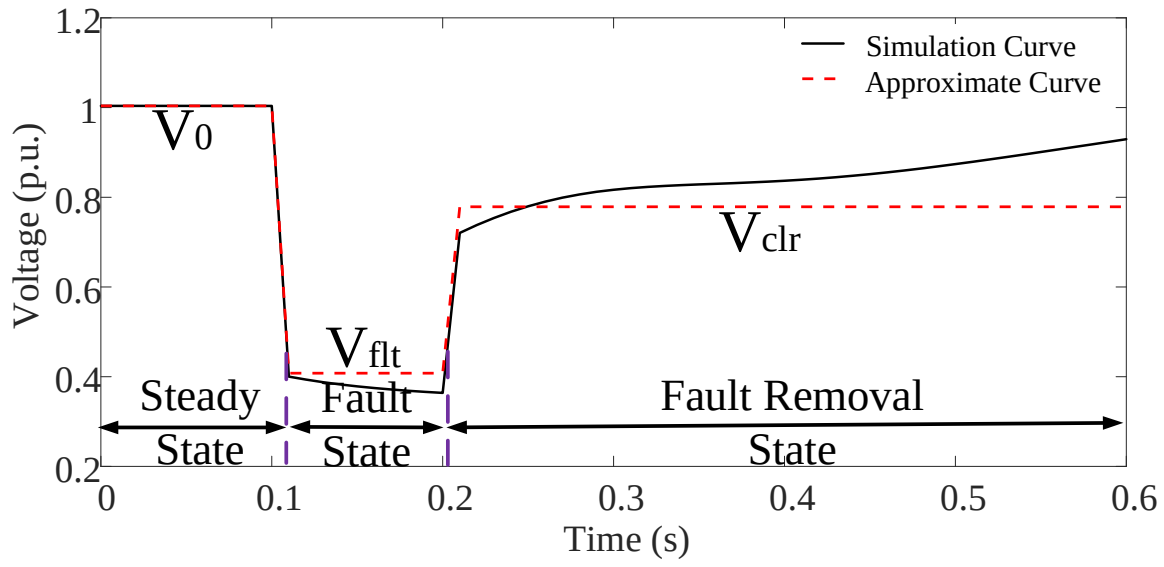

Fig. 1 'Stepped' voltage for simulating a fault effect.

The solution to the model is shown as follows, where the coefficients $A_1$, $A_2$, and $A_3$ can be calculated according to the steady-state operating state of the system and the unit parameters, as shown in Appendix A.

$$\psi'_d(t) = A_1 e^{-\frac{t}{T'_d}} + A_2 e^{-\frac{t}{T_e}} + A_3 \quad (1)$$

As shown in (1), the generator equivalent flux linkage encompasses three different components: an exponential term with $T_d'$, an exponential term with $T_e$, and a non-exponential term. $A_1$ is generally negative, indicating that the generator stator current under fault disturbance exhibits demagnetization properties, and $A_2$ is generally positive, indicating that the excitation current exhibits magnetization properties. During the fault period, due to the drop in terminal voltage, the stator current increases suddenly. However, the excitation current cannot increase immediately due to the time delay in the excitation system. Thus, the equivalent flux linkage may decrease at the beginning. After the fault is removed, the terminal voltage is restored to a higher level, and the stator current is significantly reduced. In addition, with the effect of the excitation system, the excitation current increases gradually, as does the equivalent flux linkage, which provides a support effect for system voltage recovery.

### *B. Decomposition of generators' reactive power output during the transient process*

According to the analytical expression of the flux linkage in (1), the approximate time-domain expression of the reactive power output by the generator during the transient process can be obtained, as shown in (2), where $Q_{\text{spon}}$ and $Q_{\text{exc}}$ represent the spontaneous var response and excitation control var response, respectively. $A'_1$-$A'_5$ are parameters defined in Appendix A.

$$
\begin{aligned}
&Q_g(t) = Q_{spon}(t) + Q_{exc}(t) \\
&Q_{spon}(t) = (V_q(t)\psi'_{d,spon}(t) - V_q^2(t))/x'_d - V_d^2(t)/x_q \\
&\approx V_q(t)/x'_d(A'_4 e^{-\frac{t}{T'_d}} + A'_5) - (V_q^2(t)/x'_d + V_d^2(t)/x_q) \qquad (2) \\
&Q_{exc}(t) = V_q(t)\Delta\psi'_d(t)/x'_d \\
&\approx V_q(t)/x'_d(A'_1 e^{-\frac{t}{T'_d}} + A'_2 e^{-\frac{t}{T_e}} + A'_3)
\end{aligned}
$$

The equation shows that at the moment of fault occurrence, due to flux linkage conservation and the terminal voltage sag, the spontaneous var output is much larger than that under steady state. In contrast, due to the time delay in the excitation system, the flux linkage does not change rapidly, nor does the var output excited by the excitation system control. Over time, the excitation current and the equivalent flux linkage of the winding increase gradually, and then the dynamic var support provided by the excitation system becomes prominent. It should be noted that for units with fast excitation function, the time delay is also inevitable due to the inductance effect in excitation circuits.

The peak of the spontaneous var response is mainly determined by the initial value of the flux linkage, and the initial value of the flux linkage is closely related to the generator terminal voltage setting, active power output, and reactive power output. The expression is derived in section III-C. Here we only present some discipline:

1) When the steady-state reactive power output is greater than 0, as the initial reactive power of generators increases, the initial value of the flux linkage increases, so the spontaneous reactive power response also increases.

2) For large-capacity units, the transient reactance is generally small. In this case, if the terminal voltage setting increases, the initial value of the flux linkage increases, so the spontaneous reactive power response also increases.

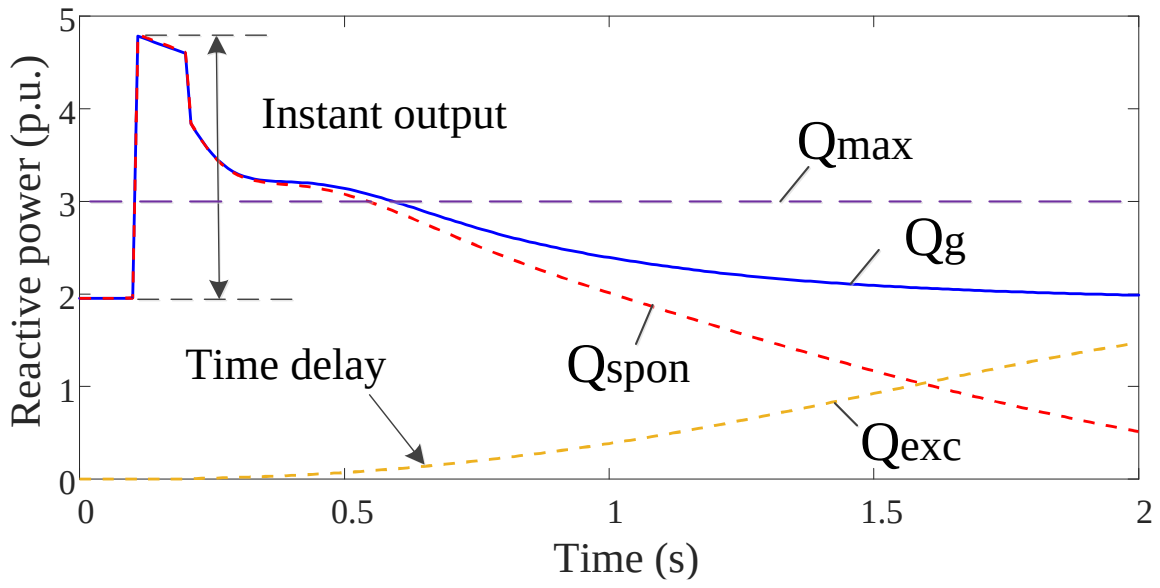

Fig. 2. Decomposition of the generator's reactive power response.

Fig 2 shows the decomposition result of the var response of a generator during the transient process. The blue line represents the var output of the generator under fault disturbance. The red line represents the spontaneous var response. The yellow line represents the excitation control var response. The purple line represents the reactive power upper limit for the device. The reactive power output of the generator increases rapidly when a fault occurs, exceeding the steady-state upper limit. The main component is the spontaneous var response at first, and the excitation control var response has a limited effect. Subsequently, the spontaneous var response declines gradually, and the effect of the excitation system control becomes prominent. This verifies the correctness of the conclusions above.

## III. Index for Voltage Support Capability During the Transient Process

### A. *Definition of the voltage support capability indexes*

The DVR for voltage stability does not consider the dynamic characteristics of the devices, so some novel indexes are introduced in this section. Since the spontaneous var response is generated due to flux linkage conservation, and the time delay in the excitation control system is reflected by the change rate of the flux linkage, it is appropriate to analyze the change in the voltage support effect from the change in the flux linkage. By eliminating the injected current in the network equations and dynamic components' flux linkage equations, it is found that the bus voltage is generated by the rotation of generators' and other dynamic components' flux linkage. Since the reactance of the high-voltage transmission lines is much larger than the resistance, and the generators' rated power angle is generally small, the x-axis component of the bus voltage is much larger than the y-axis component. Therefore, the relationship between the bus voltage and the flux linkage satisfies the superposition principle approximately, as shown in (3).

$$
\begin{aligned}
V_i(t) &= \sum_{j=1}^{N_{dyn}} V_{ij}(t) = \sum_{j=1}^{N_{dyn}} R_{ij}\psi'_{d,j}(t) \\
&\approx \sum_{j=1}^{N_{dyn}} (Z'_{xx,ij}C_{x,j} + Z'_{xy,ij}C_{y,j}) w_j \psi'_{d,j}(t)
\end{aligned} \qquad (3)
$$

where $V_{\text{ij}}(t)$ can be seen as the voltage component provided by dynamic device *j* on bus *i*. $R_{\text{ij}}$ represents the voltage support effect of 1 unit equivalent flux linkage of device *j* on bus *i*. $Z_{\text{xx}}$ and $Z_{\text{xy}}$ represent the impedance of $V_{\text{x}}$ to the *x*-axis current and the *y*-axis current, respectively. $C_{\text{x}}$ and $C_{\text{y}}$ are related to machine parameters, as shown in Appendix B. When the voltage at the moment of fault occurrence is larger, the voltage inertia support capability is more sufficient. When the bus voltage can be restored to an acceptable level faster after fault removal, the voltage recovery performance of the system is better. Therefore, if the amplitude of $V_{\text{ij}}(t)$ at the moment of fault occurrence is larger, and the area enclosed by the time axis and $V_{\text{ij}}(t)$ in a certain period is larger, the voltage support effect of device *j* is better. So, we take the value of the voltage component of generator *j* to bus *i* at the moment of fault occurrence to measure the voltage inertia, which is named the voltage inertia support capability index (VIC), and take the average value of the voltage support component of generator *j* to bus *i* in a certain period after fault removal to measure the voltage recovery performance, which is named the voltage recovery capability index (VRC). The expressions are shown in (4). It should be noted that for synchronous condensers with fast excitation systems, the time interval for VRC calculation can be counted from the moment of fault occurrence.

$$
\begin{aligned}
VIC_{ij} &= R_{ij,flt}\psi'_{d,0,j} \\
VRC_{ij} &= \frac{1}{\Delta T}\int_{T_{clr}}^{T_{clr}+\Delta T} R_{ij,clr}\psi'_{d,j}(t)dt
\end{aligned} \qquad (4)
$$

where $\triangle T$ represents the period after fault removal. The subscript "*flt*" and "*clr*" represent the variables during fault occurrence and removal. The definition of the indexes above can reflect the voltage support effect of generator *j* on bus *i* from two perspectives. They have the following characteristics.

1) The indexes are calculated based on the voltage support component under fault disturbance, which can reflect the dynamic parameters' impact on the voltage support effect. Furthermore, due to the approximate linearity between the bus voltage and the flux linkage, the voltage support effect of different generators can be summed directly.

2) Since the definition of the VRC is an integral quantity, the difference in voltage support effect at different times can be considered, and thus the performance of the excitation system can also be considered.

3) The indexes are defined based on the flux linkage. Since the flux linkage has a specific expression with respect to different variables under steady state, the system operating variables can be coordinated to reach a point that is better for STVS through the optimization for the indexes.

In summary, the indexes have considered the main factors under large disturbances, while the DVR only considers the steady-state var output. Thus, the VIC and VRC indexes are more appropriate for STVS analysis.

## B. Physical meaning of the indexes

The two voltage support capabilities measured by the VIC and VRC indexes are similar with the inertia support effect and the primary frequency modulation in the frequency response under large disturbances. The fundamental principle of two kinds of voltage support abilities is shown in Fig. 3.

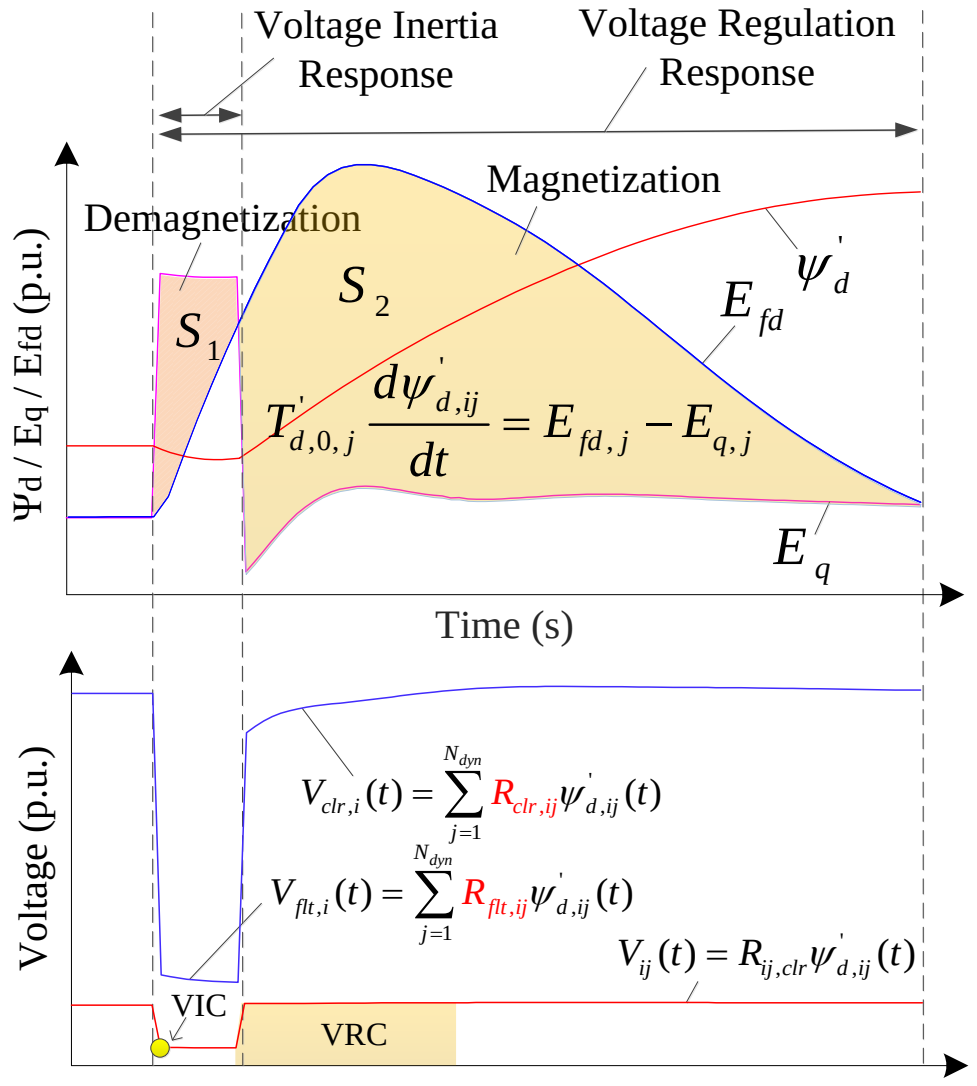

Fig. 3. Physical meaning of the VIC and VRC indexes.

For frequency response, after a large disturbance occurs, the kinetic energy of the rotating rotors in the system is instantaneously transformed into inertia support power to compensate for the unbalanced power, and the frequency gradually decreases in this period. Over time, the frequency modulation plays its role, which means that the prime mover injects power to compensate for the unbalanced power, and the inertia support power gradually decreases. In this period, the frequency drops slowly and finally reaches a turning point [27].

For voltage response, at the moment of fault occurrence, the network topology changes suddenly, and the flux linkage remains unchanged. As shown in (3), the bus voltage is approximately equal to the linear combination of the flux linkage, and the coefficients are different at different stages of the fault process. Therefore, the voltage changes suddenly. Then, the magnetic field energy inside all the machines in the system is instantaneously transformed into inertia support power, compensating for the increase in power loss of the circuit $S_1$. In this period, the flux linkage gradually decreases, and the voltage decreases as well. At the moment of fault removal, the network topology changes again, and the flux linkage remains unchanged, resulting in a sudden voltage rise and a decrease in circuit power loss. In this period, the excitation system regulation effect is obvious. The exciter injects power to compensate for the unbalanced power $S_2$ and the magnetic field energy storage, so the flux linkage begins to grow, and the voltage amplitude continues to rise. The details of other comparisons are shown in Table 1.

TABLE I COMPARISON BETWEEN FREQUENCY RESPONSE AND VOLTAGE RESPONSE UNDER LARGE DISTURBANCES

| Contrastive terms | Frequency Response $2H_{sys}\frac{d\Delta f}{dt}=\Delta P_m-\Delta P_d$ | Voltage Response $T'_{d,0,j}\frac{d\psi'_{d,ij}}{dt}=E_{fd,j}-E_{q,j}$ |
|---|---|---|
| Variables characteristics | Unique | Diverse |
| Factors to the nadir | $J$, $\Delta P_d$ , governors | $\psi$, *fault location*, excitation systems |
| Inertia definition | $J/E_{kinetic}/H$ | $\psi/E_{magnetic}$ |
| Inertia demand constraints | $d\Delta f\ /\ dt$ , $\Delta f_{nadir}$ | $V_{nadir}$ |
| Measures for improving inertia | Planning, UC | Planning, UC, $x_0$ |
| Energy conversion of inertia support | $E_{kinetic}$ → $E_{electric}$ | $E_{magnetic}$ → $E_{loss}$ |
| Energy conversion of the controller action | $E_{prime\ mover}$→$E_{electric}$+$E_{kinetic}$ | $E_{exciter}$→$E_{loss}$ + $E_{magnetic}$ |
| Key factors to the controller effect | Governor parameters | Exciter system parameters |

The above analysis indicates that the change in the magnetic field energy storage inside the machines and the network topology directly determine the change in the voltage under fault disturbance, and these two factors have been considered in the definition of the indexes in section III-A. Unlike the frequency, the bus voltage has diversity, so the evaluation for both the voltage inertia and voltage recovery capability should distinguish different buses. In addition, the main part of the voltage sag appears at the moment of fault occurrence, which is closely related to the flux linkage's initial level instead of the time constant. So, we take $\psi_{d,0}'$ rather than $T_{d,0}'$ to measure the voltage inertia. It can also be found that the voltage inertia is more controllable due to the wide value range of $\psi_{d,0}'$, while the inertia in frequency response cannot be changed by adjusting the operating point $x_0$ since w≈1.

## C. Characteristics of the indexes

Given a system and a short-circuit fault, the value of VIC is mainly determined by the initial value of stator d-axis equivalent flux linkage or magnetic field energy inside generators before the fault occurs. It is closely related to the depth of the initial voltage sag when a fault occurs. Based on the basic equations of generators, the relationship between the initial value of the flux linkage and the generator's operation variables can be derived, as shown in (5). It is found that the initial flux linkage is closely related to the initial active power output, reactive power output, terminal voltage setting of the generator. Thus, the voltage inertia provided by the generators can be changed by adjusting the operation points of the devices. For a multimachine system, the sum of the stators' d-axis equivalent flux linkage can reflect the voltage inertia of the system on a specific bus *i* when a fault disturbance occurs, and it is related to the operation status and the number of the generators, as shown in (6).

$$\psi'_{d,0} = \frac{V_{g,0}(V_{g,0}^2 + Q_{g,0}x_q) + \dfrac{(P_{g,0}^2 x_q + V_{g,0}^2 Q_{g,0} + Q_{g,0}^2 x_q)x'_d}{V_{g,0}}}{\sqrt{(P_{g,0}x_q)^2 + (V_{g,0}^2 + Q_{g,0}x_q)^2}} \tag{5}$$

$$VIC_i = \sum_{j=1}^{N_g} R_{flt,ij}\psi'_{d,0,j} \tag{6}$$

VRC mainly depends on the initial value and change rate of the generators' d-axis equivalent flux linkage after the fault is removed. It reflects the response characteristics of the generators' excitation system to voltage change under fault disturbance. According to (1), the expression of the VRC index can be converted to the form shown in (7), where △T can be assigned an appropriate value according to the requirement for voltage security in the grid code. The coefficients $A_1$, $A_2$, and $A_3$ are related to the initial operation state of the system. In addition, compared with (5), it is found that the voltage recovery capability is also closely related to the excitation system parameters. Generally speaking, if the time constant of the excitation system is smaller and the gain of the excitation system is larger, the voltage recovery performance is better, and the index value is larger as well.

$$VRC_i \approx \sum_{j=1}^{N_{dyn}} \frac{R_{clr,ij}}{\Delta T}\Big(A_{j,1}(1-\mathrm{e}^{-\Delta T/T'_{d,j}})T'_{d,j} + A_{j,2}(1-\mathrm{e}^{-\Delta T/T_{e,j}})T_{e,j} + A_{j,3}\Delta T\Big) \tag{7}$$

In addition, as shown in (4), the integral used in the definition of the VRC index can reflect the average support effect of the generator on the bus voltage during a period under fault disturbance. This average support effect essentially depends on the amount of charge passing through the generator windings with the effect of the excitation system, as shown in (8). In the period we focus on, if the charge $Q_f$ passing through the excitation winding increases and the charge $Q_d$ passing through the d-axis stator winding decreases, the voltage recovery capability of the generator is more sufficient.

$$VRC_{ij} = \frac{R_{ij}}{\Delta T}\left(x_{ad,j}Q_{f,j} - \frac{x_{ad,j}^2}{x_{f,j}}Q_{d,j}\right) \tag{8}$$

## IV. Potential Applications of the Voltage Support Capability Indexes

The voltage support capability indexes proposed in this paper can be applied in planning, ORPD, and other scenarios. When STVS is considered in these models, it is necessary to evaluate whether the voltage support ability is sufficient to maintain the secure operation of the system under fault disturbance. The security requirements for voltage under large disturbances in the grid code mainly include the requirements for the voltage sag and the voltage recovery rate. For example, for a DC station, the security requirements in China are that the voltage cannot be less than 0.85 under fault disturbance. Otherwise, there may be a risk of commutation failure. In addition, to prevent DC blocking, it is usually required that the voltage should be restored to a level higher than 0.85 within 400 ms after a fault disturbance. These requirements can be expressed in (9), which represents a voltage security region *S*. When solving the optimization models of planning or ORPD, it should be ensured that the operating point $x_0$ is within *S*. However, *S* is difficult to express explicitly. The common ways usually get the operating point through an optimization model at first. Then, time-domain simulations are conducted to obtain the voltage trajectories, and whether the point is in S is judged by comparison with the thresholds $V_{th1}$, $V_{th2}$. If it cannot meet the requirements, the sensitivity and other information should be sent back to the optimization model for iteration. Thus, this method may face problems of convergence and efficiency.

$$\begin{aligned} S_1 &= \left\{x_0 \mid \min_t\{V(t)|^{flt_i}_{x(0)=x_0}\} > V_{th,1}, \forall flt_i \in S_{flt}\right\} \\ S_2 &= \left\{x_0 \mid V(400\text{ms})|^{flt_i}_{x(0)=x_0} > V_{th,2}, \forall flt_i \in S_{flt}\right\} \end{aligned} \tag{9}$$

$$\begin{aligned} S_1 &= \{x_0 \mid VIC_i(x_0, flt_i, Y) > VIR_{flt_i}, flt_i \in S_{flt}\} \\ S_2 &= \{x_0 \mid VRC_i(x_0, flt_i, Y) > VRR_{flt_i}, flt_i \in S_{flt}\} \end{aligned} \tag{10}$$

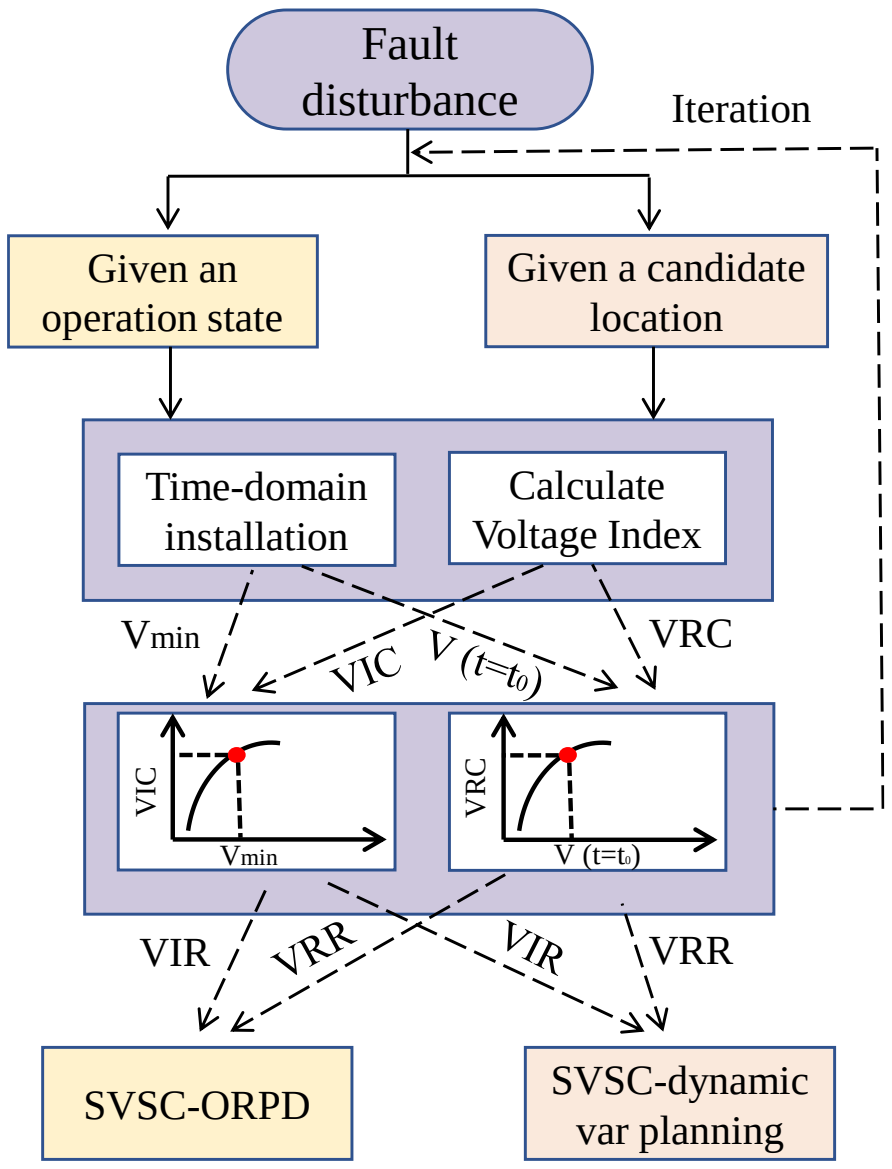

Fig. 4. Applications of VIC/VRC in planning/ORPD.

Through the indexes proposed in this paper, $S_1$ and $S_2$ can be expressed as (10), which are the same sets as (9). $VIC_i$ and $VRC_i$ in (10) can be calculated based on the operating point $x_0$, network parameters *Y*, and an anticipated fault $flt_i$, which are shown in (5)-(7). *VIR* and *VRR* can be obtained through time-domain simulations. However, once the thresholds are determined, they do not depend on the operating status of devices. Therefore, (10) can be added as constraints to the models of planning or ORPD directly, as shown in Fig.4. With these constraints, the optimization models can be solved without iterations with time-domain simulation, which means they are decoupled.

The following processes can be adopted for obtaining the security thresholds. For a given anticipated fault, by adjusting the operation status of generators and other var devices in the same zone of the fault location, the var output of different devices is coordinated to maintain the regional voltage at an acceptable level. Then, time-domain simulations are conducted to obtain the voltage trajectories. The voltage nadir after the fault occurs and the voltage amplitude at a specific moment are recorded, and the corresponding VIC and VRC indexes are calculated at the same time. Finally, the corresponding $V_{nadir}$-VIC curve and $V_{t=t0}$-VRC curve can be obtained. By selecting the critical points according to the requirements of the grid code, the corresponding VIC/VRC values of the critical points are the requirements. Fig. 5 shows an example for evaluating the VRR under a specific fault. It should be noted that the voltage support requirement is closely related to the load characteristics, so the evaluation processes should be conducted periodically for online applications.

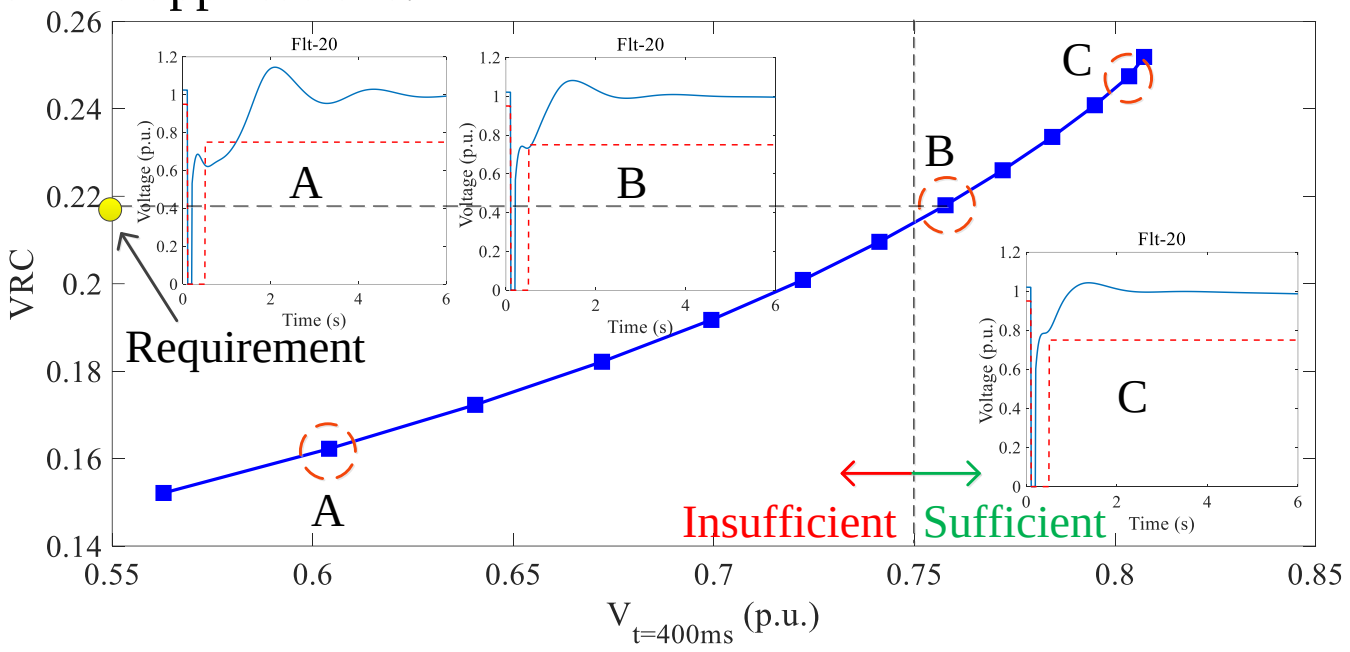

Fig. 5. Assessment for the requirement of voltage recovery capability during the transient process under a specific fault.

## V. Case Study

In this section, we will conduct time-domain simulations and calculate the indexes based on the IEEE 39-bus system.

### A. Verification of the expressions of the generators' flux linkage and bus voltage

Based on time-domain simulations and the expression in (3), the voltage support components of different rotating electrical machines in the system under a fault disturbance are shown in Fig. 6. We can find that when the voltage support effect of the synchronous generators' and induction motors' flux linkage are superimposed, the actual voltage curves are obtained. This verifies the correctness of the relationship between the flux linkage and the bus voltage in (3).

When the proportion of induction motors in the load is 30%, 60%, and 90%, the support components of generators and the actual bus voltage are denoted with red, blue, and green lines, respectively. We can find that as the proportion of induction motors increases, the proportion of synchronous generators' support components in the bus voltage decreases continuously, and the impact of induction motors' dynamic characteristics on the bus voltage becomes increasingly obvious. Since the induction motors do not have excitation systems, their winding flux linkage can only follow the change in their terminal voltage, and the STVS of the system becomes worse.

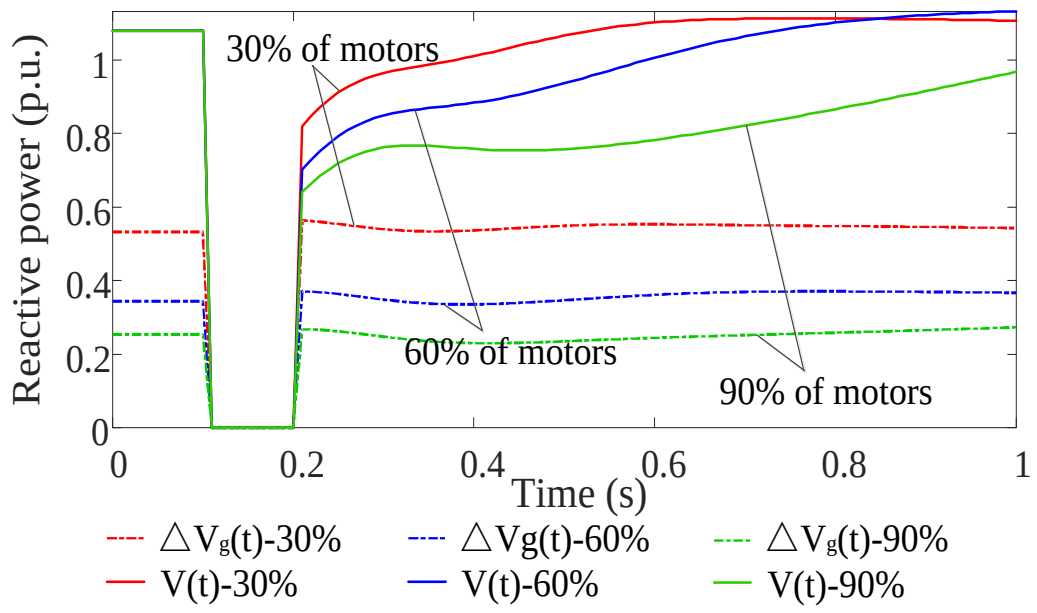

Fig. 6. Voltage support components of all generators with different proportions of induction motors.

Since the time-domain expression of the flux linkage is derived by assuming that the terminal voltage curve exhibits a two-stage 'stepped' shape, we will analyze the errors caused by this assumption. The comparison between the simulation curve and the calculated curve of the generator's flux linkage is shown in Fig. 7, in which the red line is the calculation result, and the blue line is the simulation result. It is found that the calculated curve is slightly larger than the simulation curve during the fault period. The reason is that the generator terminal voltage exhibits a downward trend in this period, which causes the stator current to increase slightly, and thus the demagnetization effect of the current is stronger. As a result, the actual curve is smaller than the calculated curve. Since the terminal voltage exhibits an upward trend and the deviation between the voltage setting and the current value decreases after fault removal, the actual growth rate of the flux linkage is slightly slower than the calculation result. It can be seen that the maximum error during the fault period is approximately 0.39%, and the maximum error within 400 ms is approximately 1.17%, which is acceptable for fast estimation and mechanism analysis.

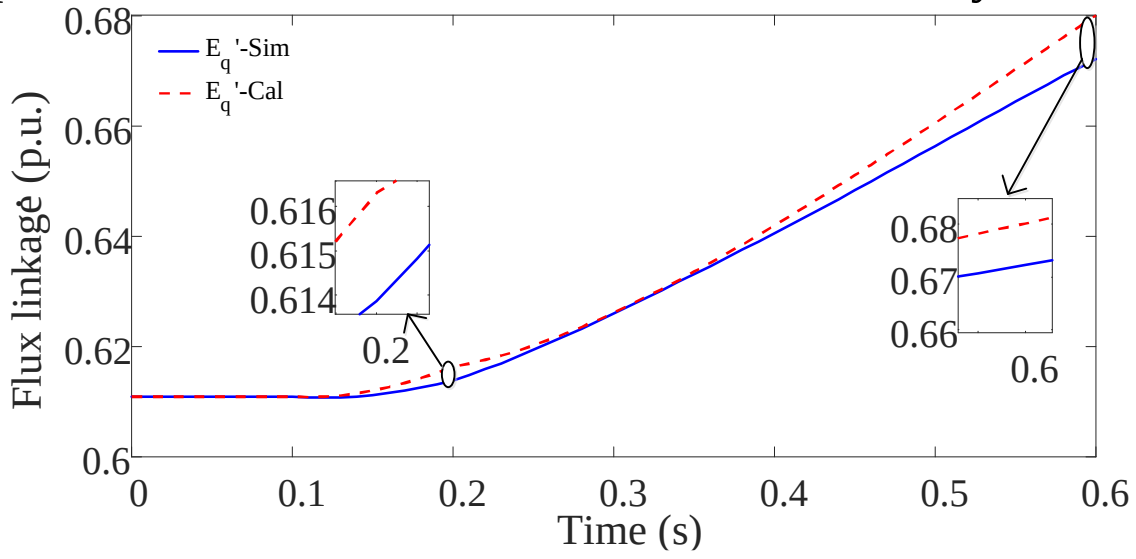

Fig. 7. The calculated and simulation curves of the flux linkage.

### B. Impact of installing var devices on voltage support ability

When the voltage inertia of the system is insufficient, the bus voltage sag under fault disturbance may have a bad impact on the system, such as DC commutation failure. In this case, to improve the STVS of the system under fault disturbance, some dynamic var devices can be installed. It is assumed that a three-phase short-circuit fault occurs at the terminal of line 17-27, and the fault duration is set as 0.1 s. The voltage trajectory of bus 15 is investigated, and the security threshold is assumed to be 0.75. Under the initial operation state, the voltage sag of bus 15 is 0.53 when the fault occurs, indicating that the voltage inertia is insufficient. By installing a synchronous condenser on different buses respectively, the corresponding values of the VIC index and VRC index are calculated, as shown in Fig.8 (a), and the voltage trajectories of bus 15 under different cases are analyzed, as shown in Fig.8 (b).

As shown in Fig.8 (a), as the electric distance between the candidate bus and bus 15 decreases, the voltage inertia support from the condenser on bus 15 becomes more sufficient. When VIC is larger than 0.58, the voltage sag of bus 15 is able to meet the security constraint $V_{min}>V_{th}$. Thus, it can meet the requirement when the condenser is installed on bus 15 or 16. For bus 15, the system voltage inertia can be increased to 0.70 after the condenser is installed on it, and the voltage nadir is increased to 0.84. The support effect of the condenser on bus 15 is the most sufficient, which is consistent with our experience. From Fig.8 (a) and (b), we can find that with the installation of synchronous condensers, the voltage recovery performance is also improved to a certain extent, and the improvement is similar to that of the voltage inertia. This is because when the voltage inertia increases, the voltage nadir improves, and the voltage can recover to the initial level faster.

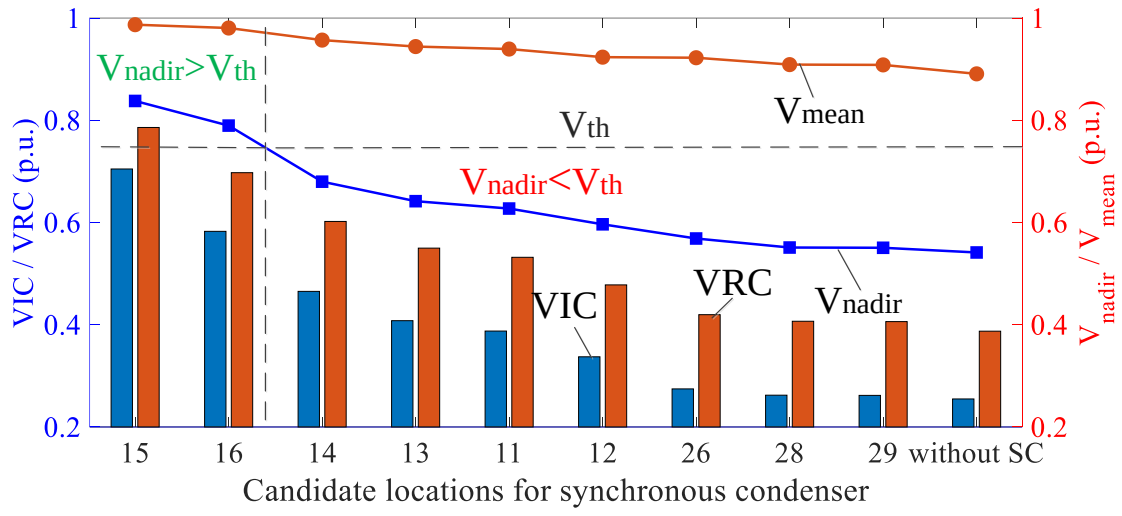

(a) Voltage indexes and voltage amplitude of bus 15.

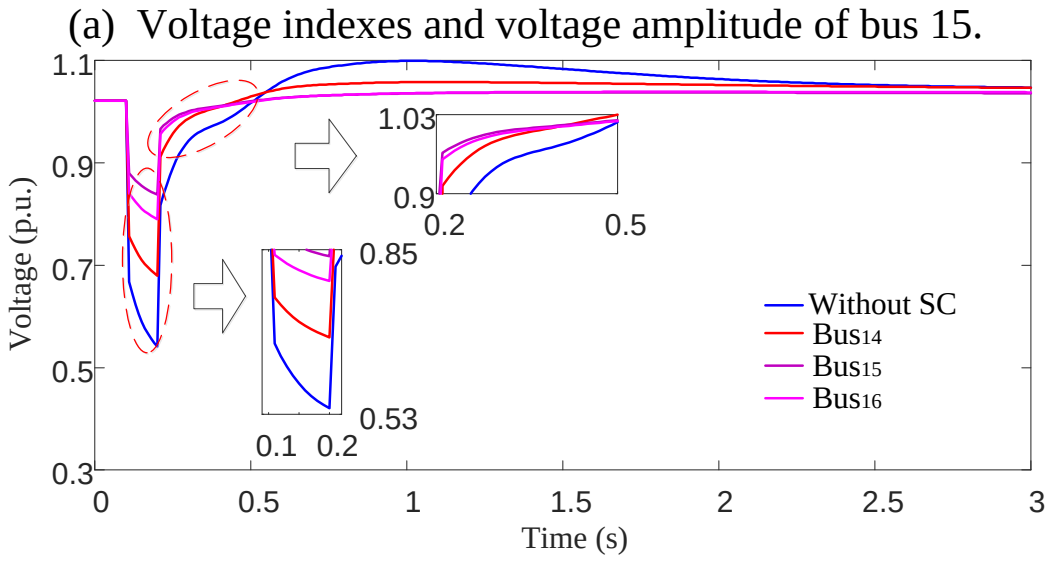

(b) Voltage trajectories of bus 15.

Fig. 8. Impact of the condenser installed on different buses.

## C. *Impact of operating points on voltage inertia*

The voltage inertia support of the var devices in operation can be fully exploited by coordinating their var output, terminal voltage settings, etc. A large number of operating points are generated randomly, and then time-domain simulations and the VIC index are calculated. The results are shown in Fig.9 (a). The voltage inertia and the voltage nadir show an approximately linear relationship. This is because the VIC index has considered the nonlinearity between the steady-state operating variables and the voltage sag. When the requirements for voltage nadir are given, the corresponding VIR can be obtained. Three operating points A, B, and C are selected for further study. The voltage amplitude of bus 15 under the fault occurrence is the largest under C, and it is improved by about 0.1 compared to that under A, which shows the voltage inertia of the system has been greatly improved. The VIC index under C also increases by 0.11 compared to that under A, which verifies the effectiveness of the VIC index. It should be noted that the voltage sag improvement by adjusting the operating point cannot achieve the same performance as installing SC.

The settings under A, B, and C are shown in Fig.9 (b), in which the areas in different colors represent different initial values of the flux linkage or magnetic field energy, and the points in red, blue, and green correspond to operating state A, B, and C. The magnetic field energy will increase as the terminal voltage or the var output improves. The var output of most generators is at a low level under A, which means that A pursues more DVR under steady state. In contrast, the var output and terminal voltage settings of most generators are at a higher level under B and C. Moreover, their initial value of the flux linkage is larger, which means that B and C pursue more system voltage inertia. The system voltage inertia under C is the most sufficient among the three operation states, so the voltage amplitude is the highest among the three operating points.

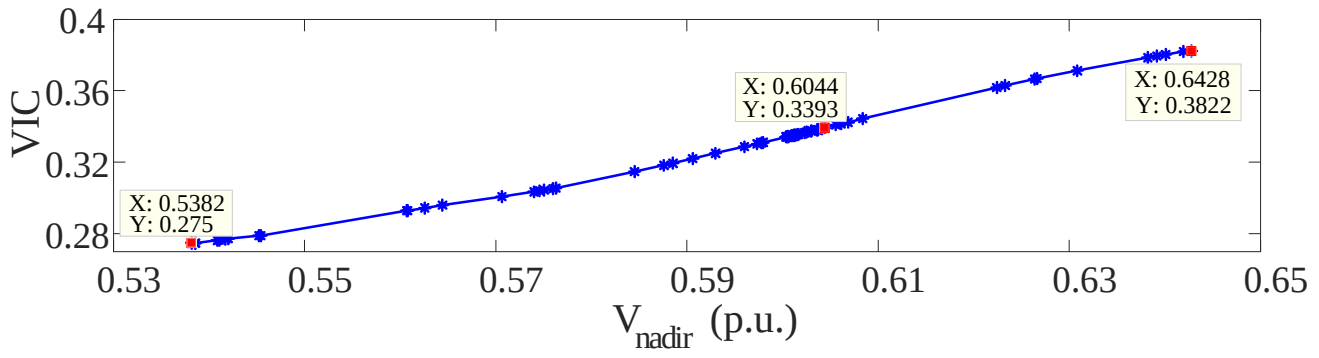

(a) Voltage nadir at the terminal of the fault line and the VIC index.

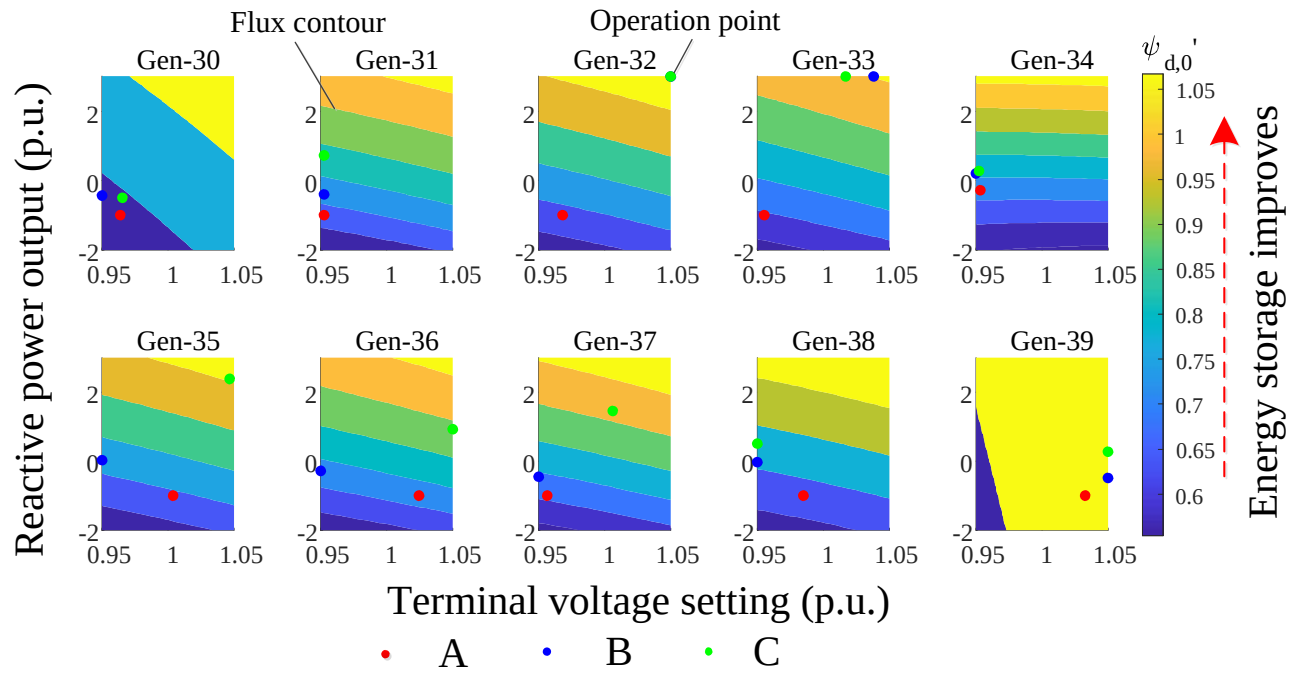

(b) Settings of different generators and the corresponding initial flux linkage.

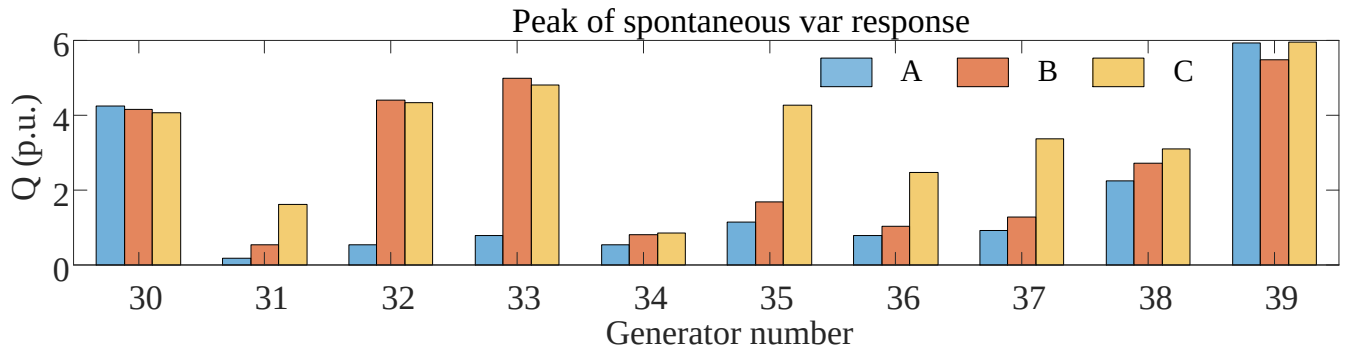

(c) Spontaneous var output of generators.

Fig. 9. Comparisons of three different operation points.

Fig.9 (c) shows the spontaneous var output of all generators after the fault occurs. For most generators, as the initial value of the flux linkage increases, the dynamic var that the generators can provide when the fault occurs is more sufficient. The reason is that the flux linkage cannot increase immediately when the fault occurs, and the var support that the generator can provide depends on the difference between the flux linkage and the terminal voltage. The greater the difference is, the greater the dynamic var output. For example, the spontaneous var output of 32, 33 under A is much smaller compared with that under B and C. From Fig.9 (b), we find that the flux linkage under A is smaller than that under B and C. Since the differences between the flux linkage and the terminal voltage of generator 31, 32 are the smallest under A, their spontaneous var output is the smallest among the three points. Thus, it is essential to consider the support effect of the initial magnetic field energy instead of DVR.

## D. *Impact of operation points on voltage recovery ability*

To analyze the impact of changing the operation point on the voltage recovery ability under different faults, 8 typical anticipated faults are generated. The fault locations are set at the terminals of each line respectively, the duration is set as 0.1 s, and the voltage threshold is set as 0.75. First, the voltage support requirements under different faults are evaluated. According to the procedure in section IV, the requirements for different faults are obtained, which are shown as yellow dots in Fig.10 (a). The blue bars in the figure represent the VRC index under operation state A. It can be seen from the figure that under A, the voltage support capability of the system cannot meet the requirements of all the faults. By adjusting the operation status of the reactive power devices, two different operating states B and C are obtained, as shown in the orange and yellow bars in Fig.10 (a). It is found that when the anticipated faults occur under B, the voltage support capability of the system can meet the requirements of some faults. In contrast, the voltage support capability of the system under C can meet all fault requirements.

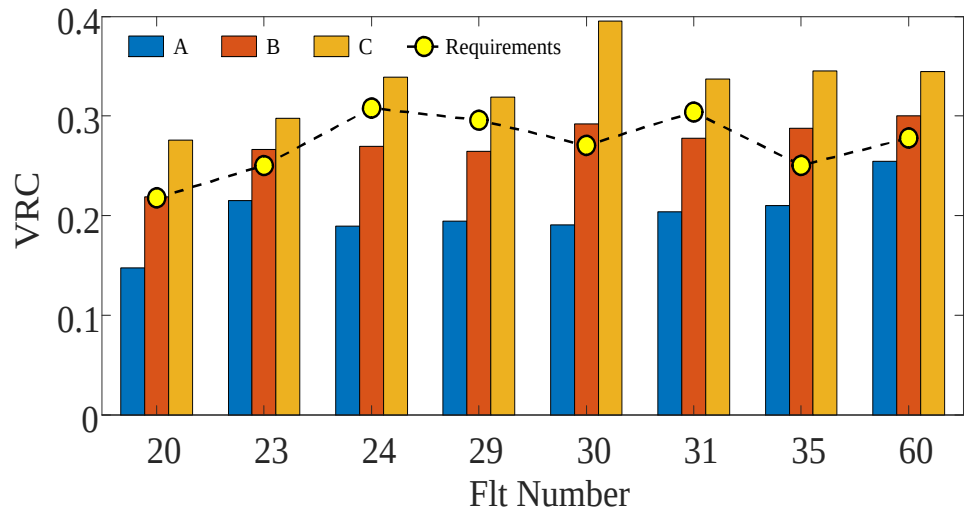

(a) VRC index and the requirements.

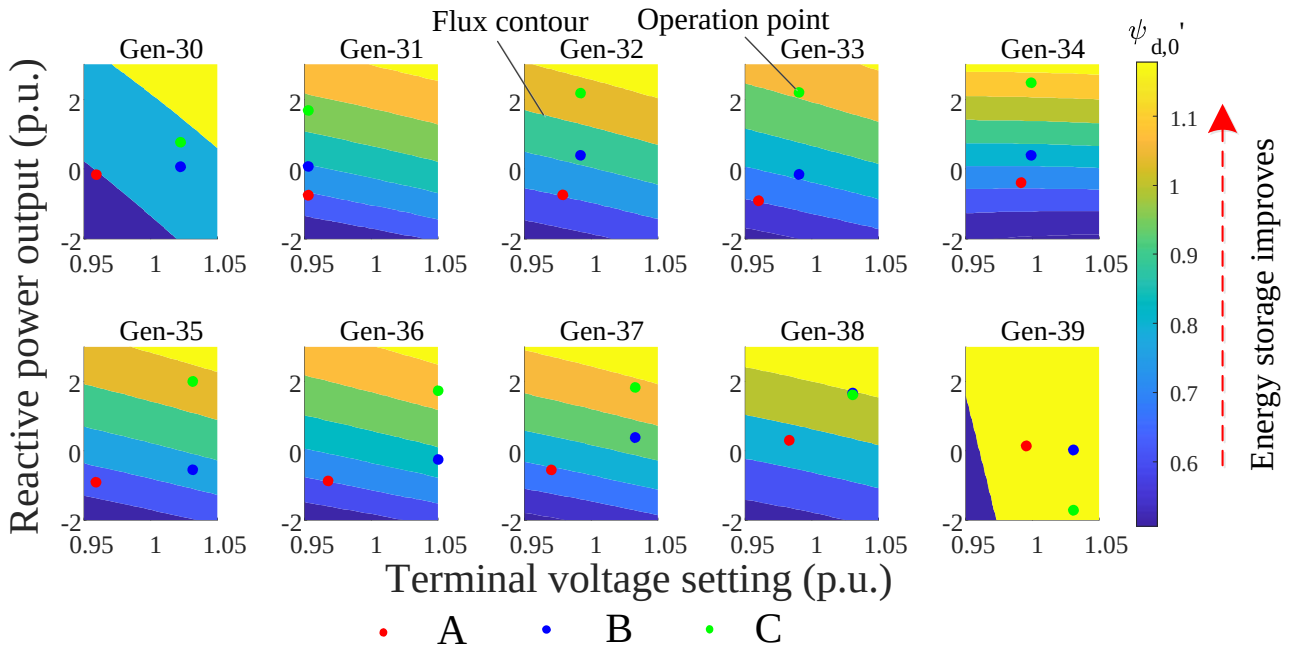

(b) Settings of different generators and the corresponding initial flux linkage.

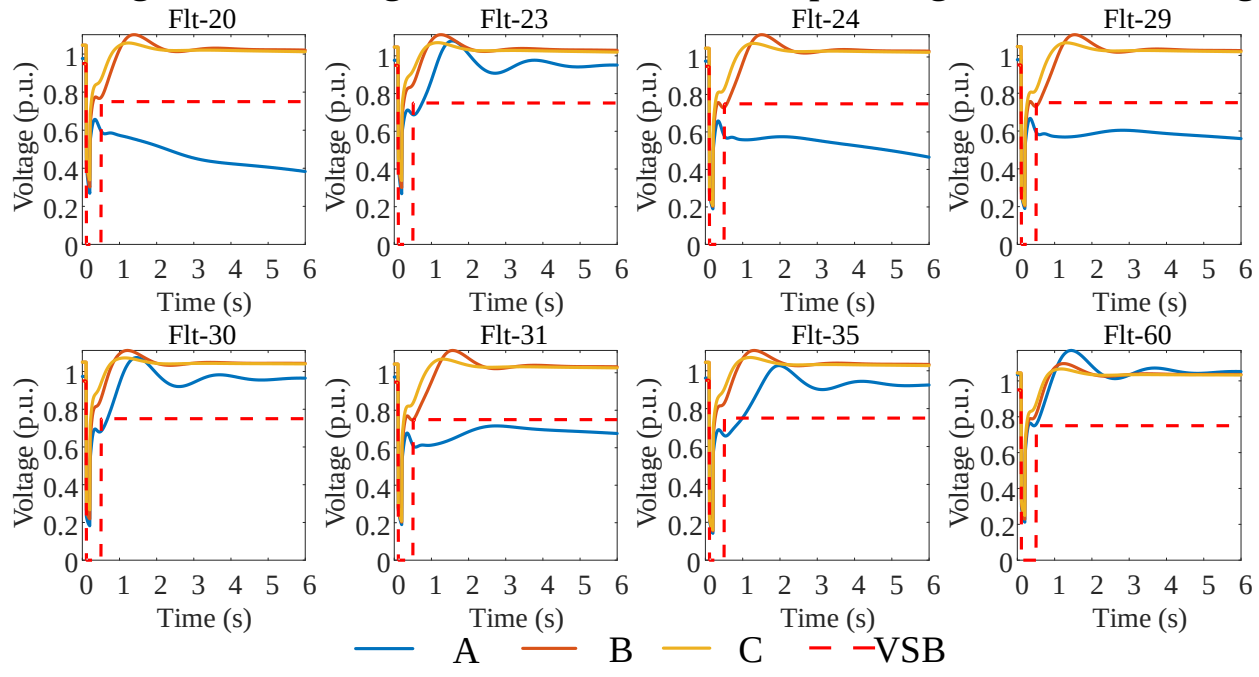

(c) Voltage trajectories.

Fig. 10. Comparisons under three different operating statuses.

As shown in Fig.10 (b), compared with operating point A, the terminal voltage settings of most units under B and C are at a higher level. In general, the increase in terminal voltage setting can increase the generator d-axis equivalent flux linkage, and thus the voltage support ability of the system is improved. Therefore, compared with operating point A, operating point B can meet the requirements of some faults. However, some faults' voltage support requirements cannot be met under B. The reason is that the flux linkage under B is not large enough, and the voltage support ability of the system has not been fully utilized. In contrast, it can be seen from Fig.10 (b) that under C, the terminal voltage and reactive power output of most units in the system are at the highest level, and the average level of initial var output is nearly 190 MVar larger than that under B. Considering the bus voltage upper limit, it is necessary to switch off some capacitors and switch on some reactors under operating point C.

Fig.10 (c) shows the voltage trajectories under different faults, in which the blue, red, and yellow lines represent operation points A, B, and C. The red dotted lines are the voltage security boundary (VSB) required by the grid code. When the voltage trajectory is located above the red dotted line, it indicates that the voltage security constraints can be met. It can be seen that under operating point A, the occurrence of faults 23, 30, 35, and 60 causes the FIDVR, and the occurrence of faults 20, 24, 29, and 31 causes short-term voltage instability. For operating point B, it can be seen that most of the system voltage trajectories under fault disturbance are located near the security boundary. Under operating point C, the voltage trajectories under each fault are located above VSB, which is consistent with the evaluation results of the VRC index shown in Fig.10 (a). It indicates that the thresholds of the VRC index can quantify the STVS requirements of the grid code accurately, which verifies the effectiveness of the index.

## VI. Conclusion

This paper reveals that the dynamic var support provided by synchronous generators and condensers under fault disturbance includes two parts: spontaneous var response and excitation control var response. The former is generated at the moment of fault occurrence, while the latter has a time delay. Further studies show that the voltage support effect of the generator is essentially determined by its internal magnetic field energy and the network coupling degree, and the change in the magnetic field energy during the transient process is affected by the difference between the loop's active power loss and the exciter injection power. Based on this, two novel indexes for measuring the system's voltage inertia support capability at the moment of fault occurrence and the voltage recovery performance after fault clearance are proposed, respectively. It has been found that the two capabilities are affected by the system operation state, such as the operating points and the on-off status of the units. In addition, the excitation system parameters (gain, time constant) only have an impact on the voltage recovery performance instead of the voltage inertia. Finally, the potential applications of the indexes are introduced in the scenarios of planning and ORPD, in which the thresholds of the indexes can be used as the security constraints in the optimization models. Several case studies show that the indexes proposed in this paper can reflect the changes in the voltage support capability of the system correctly, and the evaluation results can quantify different fault requirements for voltage support capability accurately, which proves the correctness and effectiveness of the analysis method and the indexes. Future study can focus on exploiting the voltage inertia of power electronic devices.

## APPENDIX

### A. Parameters in flux linkage expression

$$\begin{cases} A_1 = \dfrac{T_e(\psi'_{d,0} - V_{q,flt})x_d^2 + T'_{d0}(E_{fd,0} - V_{q,flt} + K_A(V_0 - V_{flt}))x_d'^2 + (-\psi'_{d,0}T'_{d0} - E_{fd,0}T_e + (T'_{d0} + T_e)V_{q,flt})x_d x'_d}{x_d(T_e x_d - T'_{d0}x'_d)} \\ A_2 = T_e x'_d K_A(V_0 - V_{flt}) / (T'_{d0}x'_d - T_e x_d) \\ A_3 = (V_{q,flt}(x_d - x'_d) + (E_{fd,0} + K_A(V_0 - V_{flt}))x'_d) / x_d \\ A'_1 = K_A T'_{d,0}(V_0 - V_{flt})x_d'^2 / (x_d(T_e x_d - T'_{d,0}x'_d)) \\ A'_2 = -K_A T_e(V_0 - V_{flt})x'_d / (T_e x_d - T'_{d,0}x'_d) \\ A'_3 = K_A(V_0 - V_{flt})x'_d / x_d \\ A'_4 = E_{q,0} + (x'_d - x_d)V_{q,flt} / x_d - E_{fd,0}x'_d / x_d \\ A'_5 = V_{q,flt} + (E_{fd,0} - V_{q,flt})x'_d / x_d \end{cases} \quad (11)$$

### B. Parameters of generators and induction motors

$$\begin{cases} G_x = (x_q - x_d)\sin\delta\cos\delta / (x_q x_d) \quad , C_x = \sin\delta / x'_d \\ G_y = -(x_q - x_d)\sin\delta\cos\delta / (x_q x_d), C_y = -\cos\delta / x'_d \\ B_x = (x_d + (x_q - x'_d)\sin^2\delta) / (x_q x'_d) \\ B_y = -(x'_d + (x_q - x'_d)\cos^2\delta) / (x_q x'_d) \\ K_Z = -C_I V_{LD,0}^2 / (P_m X'), \quad X = X_1 + X_\mu \\ C = \dfrac{-(X - X')/X'}{-1 - (X - X')/X' - j2\pi f_0 s_0 T'_0} = C_R + jC_I \\ X_\mu = 2\pi f_0 T'_0 R_2 - X_2, \quad X' = X_1 + X_2 X_\mu / (X_2 + X_\mu) \end{cases} \quad (12)$$